\documentclass[9pt,twocolumn,twoside]{optica_edited}
\setboolean{shortarticle}{true}

\usepackage{siunitx}
\usepackage[version=3]{mhchem}		%subscripts chemical
\usepackage{nicefrac}
\usepackage{threeparttable}
\usepackage{makecell}
\usepackage{multicol}
\title{Carbon dioxide absorption spectroscopy with \\a mid-infrared silicon photonic waveguide}

\author[1*]{Floria Ottonello-Briano}
\author[1]{Carlos Errando-Herranz}
\author[2]{Henrik R\"odjeg\aa rd}
\author[2]{\\Hans Martin}
\author[1]{Hans Sohlstr\"om}
\author[1]{Kristinn B. Gylfason}

\affil[1]{Micro and Nanosystems, KTH Royal Institute of Technology, Malvinas v\"ag 10, 10044 Stockholm, Sweden}
\affil[2]{Senseair AB, Stationsgatan 12, 82471 Delsbo, Sweden}

\affil[*]{Corresponding author: floria@kth.se}

\begin{abstract}
Carbon dioxide is a vital gas for life on Earth, a waste product of human activities, and widely used in agriculture and industry. Its accurate sensing is therefore of great interest.
Optical sensors exploiting the mid-infrared light absorption of \ce{CO2} provide high selectivity, but their large size and high cost limit their use. Here, we demonstrate \ce{CO2} gas sensing at \SI{4.2}{\micro\meter} wavelength using an integrated silicon waveguide, featuring a sensitivity to \ce{CO2} of \SI{44}{\percent} that of free-space sensing. The suspended waveguide is fabricated on a silicon-on-insulator substrate by a single-lithography-step process, and we route it into a mid-infrared photonic circuit for on-chip-referenced gas measurements. Its demonstrated performance and its simple and scalable fabrication make our waveguide ideal for integration in miniaturized \ce{CO2} sensors for distributed environmental monitoring, personal safety, medical, and high-volume consumer applications.
\end{abstract}

\begin{document}
\sisetup{separate-uncertainty}

\maketitle
\thispagestyle{fancy}

Carbon dioxide (\ce{CO2}) is an atmospheric trace gas and, being the carbon source in the carbon cycle, it is vital to life on Earth. It is also a waste product of human activities and massively used in agriculture and industry. The atmospheric \ce{CO2} concentration is growing at an ever increasing rate and reached \SI{410}{ppm} in 2018~\cite{Keeling2017scripps}. Besides affecting Earth's climate~\cite{Lacis2010atmospheric,Matthews2009proportionality}, elevated \ce{CO2} levels increase air pollution mortality~\cite{Jacobson2008causal}, and gross leakage of \ce{CO2} puts personnel at risk of asphyxiation~\cite{Guais2011toxicity}. Indoors, high \ce{CO2} levels deteriorate human cognitive function and decision-making~\cite{Satish2012co2,Allen2016associations,Allen2018airplane}, with consequences spanning from reduced attention and productivity in classrooms and offices~\cite{Satish2012co2,Allen2016associations} to an increased risk for car and airplane accidents~\cite{Allen2018airplane}. Extensive and accurate sensing of \ce{CO2} is therefore crucial.

\begin{figure}[b!]
	\centering\includegraphics[width=\columnwidth]{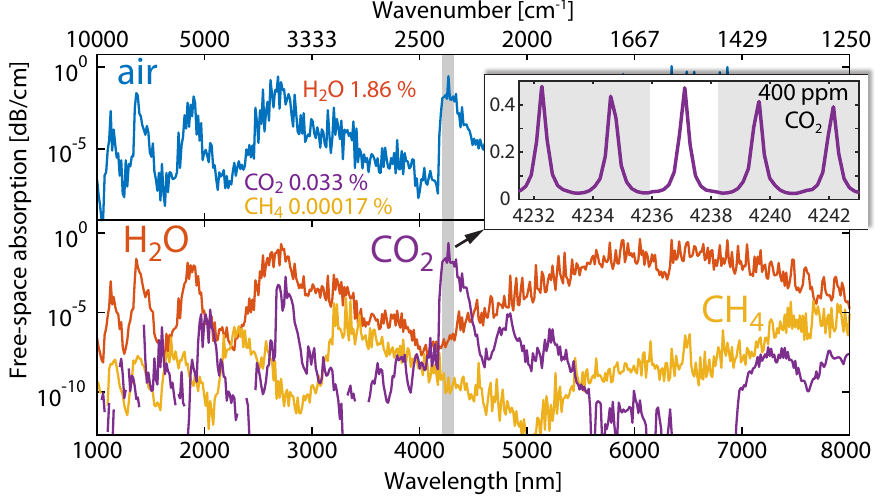}
	\caption{Absorption spectra of typical air at \SI{296}{\kelvin} and \SI{1}{atm} and its mid-IR-active constituents: water vapor, carbon dioxide, and methane~\cite{Gordon2017hitran2016}. Inset: zoom-in of five peaks in the fundamental asymmetric stretching band of \ce{CO2}. Highlighted, the range we employ for absorption spectroscopy with our waveguide.}
	%	\caption{Absorption spectra of typical air at \SI{296}{\kelvin} and \SI{1}{atm} and its mid-IR-active constituents: water vapor (\SI{18600}{ppm}), carbon dioxide (\SI{330}{ppm}), and methane (\SI{1.7}{ppm})~\cite{Gordon2017hitran2016}. Inset: Zoom-in of five peaks in the fundamental asymmetric stretching band of \ce{CO2}. Highlighted, is the wavelength range we employ for \ce{CO2} absorption spectroscopy with our waveguide.}
	\label{fig:atm_absorpt}
\end{figure}
\begin{figure*}[ht]
	\centering\includegraphics{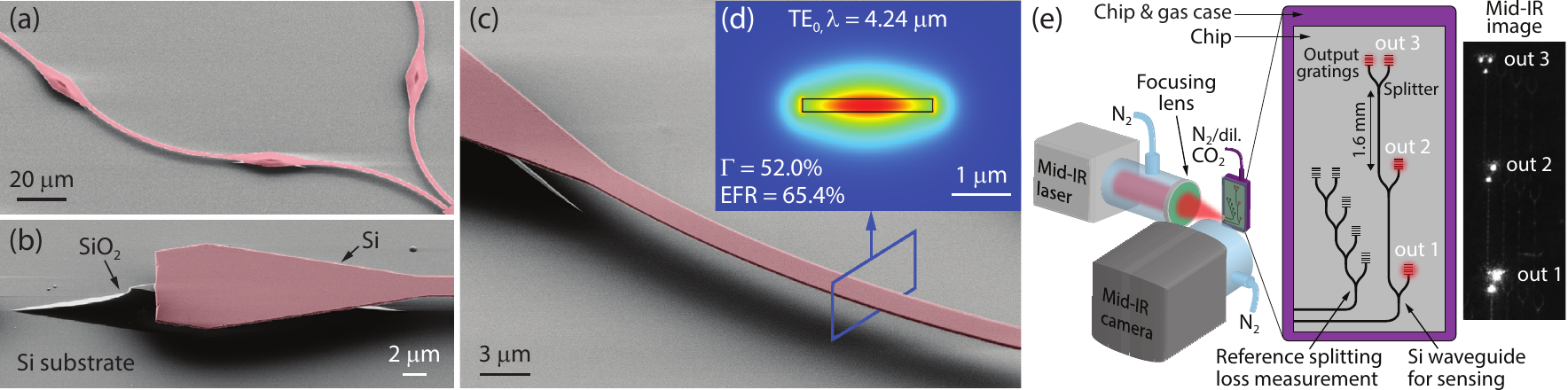}
	\caption{(a), (b), (c)~False-color SEM images of the fabricated \ce{Si} photonic waveguide. (a) Waveguides branching out from a splitter. (b) A support structure where the \ce{Si} waveguide was partially removed to expose the \ce{SiO2} pillar. (c) A close-up view of the suspended waveguide. (d)~Cross-sectional FEM simulation of the waveguide, displaying the electric field profile of the de-confined fundamental TE mode. (e)~Schematic of the setup used for \ce{CO2} sensing, including a sample mid-IR image of the operating waveguide.}
	\label{fig:device_setup}
\end{figure*}
Optical \ce{CO2} sensors would benefit most applications, due to their high selectivity, fast response, and minimal drift, compared to electrochemical and metal-oxide semiconductor-based sensors~\cite{Hodgkinson2013optical}. However, the adoption of traditional non-dispersive infrared \ce{CO2} sensors, with a free-space configuration, is limited by their large size, high cost, and high power consumption.
In contrast, optical \ce{CO2} sensors based on integrated photonic waveguides, which allow a light path as long as tens of centimeters to fit in a volume smaller than a few cubic millimeters, could achieve the level of miniaturization and power consumption required for mobile applications.

Despite the advantages of miniaturized optical absorption \ce{CO2} sensors, their development has been hindered by the lack of suitable optical components for the mid-infrared (mid-IR) spectral range, where the optical sensing of ambient \ce{CO2} with high selectivity and sensitivity is optimal. At wavelengths around \SI{4.26}{\micro\meter}, \ce{CO2} presents strong isolated absorption peaks that do not overlap with those of other gases commonly present in ambient air, such as water vapor, as shown in Fig.~\ref{fig:atm_absorpt}. 
The recent progress in light sources~\cite{Jung2017nextgeneration,Lin2017midinfrared,Fedeli2018midinfrared}, detectors~\cite{Tittl2015switchable,Cheng2016progress,Lin2017midinfrared,Hu2017silicon,Deckoff-Jones2018chalcogenide,Fedeli2018midinfrared,Huang2018waveguideintegrated}, and integrated waveguides~\cite{Lin2017midinfrared,Hu2017silicon,Fedeli2018midinfrared,Zou2018midinfrared} for the mid-IR is now accelerating the development of on-chip optical \ce{CO2} sensors.  

To be the core element of miniaturized optical \ce{CO2} sensors, photonic waveguides must fulfill two key requirements. They must support light modes at \SI{4.2}{\micro\meter} wavelength with a large portion of the field outside the waveguide core material, to enable interaction with the gas, and they should have a low base propagation loss, as this limits the applicable waveguide length.
%The ratio between these characteristics determines the attainable limit of detection, i.e.\ sensing performance, of the waveguide, and is expressed by the figure of merit $ \mathrm{FOM}\mkern-1mu =\mkern-1mu \nicefrac{\Gamma}{\alpha} $~\cite{Kita2018are}.
The ratio between these characteristics fully determines the sensing performance of the waveguide, and is expressed by the figure of merit $ \mathrm{FOM}\mkern-1mu =\mkern-1mu \nicefrac{\Gamma}{\alpha} $, introduced by Kita et al.~\cite{Kita2018are}.
Here, $ \alpha $ is the waveguide base attenuation coefficient, including all losses not due to \ce{CO2} absorption, such as material, scattering, curvature, and substrate losses. $ \Gamma\mkern-1mu =\mkern-1mu \nicefrac{\partial n_{\mathrm{mode}}}{\partial n_{\mathrm{clad}}} $, where $n_{\mathrm{mode}}$ is the light mode's effective index, is the external confinement factor expressing the waveguide's sensitivity to changes in the cladding's refractive index $n_{\mathrm{clad}}$~\cite{Veldhuis2000sensitivity,Robinson2008firstprinciple,Kita2018are}.
Contrarily to the evanescent field ratio (EFR), i.e.\ the portion of optical power propagating outside the waveguide core, $ \Gamma $ correctly describes the sensitivity of any waveguide, including those with high core-cladding refractive index contrast~\cite{Robinson2008firstprinciple}.
%$ \Gamma $ differs from the evanescent field ratio (EFR), i.e.\ the portion of optical power propagating outside the waveguide core, especially for waveguides with high core-cladding refractive index contrast~\cite{Robinson2008firstprinciple}.
The FOM, in conjunction with the waveguide length and the signal-to-noise ratio of the measurement setup, determines the achievable limit of detection of the system.

A variety of integrated waveguides for the mid-IR have been presented~\cite{Lin2017midinfrared,Hu2017silicon,Fedeli2018midinfrared,Zou2018midinfrared}. A particularly attractive waveguide material is silicon (\ce{Si}), because it combines a large transparency window in the mid-IR, wide availability, and a well-established mass production infrastructure. \ce{Si}-waveguide-based sensing of methane has been demonstrated at \SI{1.65}{\micro\meter} wavelength~\cite{Tombez2017methane}. Sensing of \ce{CO2} at \SI{4.2}{\micro\meter}, however, remains challenging, because the commonly used silicon dioxide (\ce{SiO2}) cladding is optically absorbing at that wavelength. 
%silicon dioxide (\ce{SiO2}) is optically absorbing, and thus not suitable to be used as cladding material, as is usually done in near-IR silicon photonics.
%Siebert and M\"uller~\cite{Siebert2005infrareda,Siebert2005infraredb} presented an analysis of the key parameters involved in the design of mid-IR waveguides for \ce{CO2} sensing, and proposed silicon (\ce{Si}) waveguides carved out of bulk \ce{Si} with a process requiring two hard masks, oxidation, and three different etching steps. Their waveguides have a high vertical aspect ratio, thus a large etched sidewall surface, and consequently high scattering loss. Furthermore, the highest EFR achievable by their optimal design is only \SI{24}{\percent}. 
Siebert and M\"uller~\cite{Siebert2005infrareda} proposed carving \ce{Si} waveguides for \ce{CO2} sensing out of bulk \ce{Si}. However, their fabrication process is complex, and their waveguide design leads to a large etched sidewall surface, resulting in high scattering loss. Moreover, the highest EFR achievable is only \SI{24}{\percent}.
More recently, Ranacher et al.\ showed \ce{CO2} sensing with a poly-crystalline \ce{Si} strip waveguide on an \ce{SiO2} cladding, with a simulated EFR of \SI{7.4}{\percent}~\cite{Ranacher2018characterization}, and the same waveguide on a partially suspended silicon nitride membrane on \ce{SiO2}-on-\ce{Si} support structures, formed by through-wafer back-side etching, with a simulated EFR of \SI{10.7}{\percent}~\cite{Ranacher2018midinfrared}. 

Here, we present a partially suspended mid-IR \ce{Si} waveguide with a high external confinement factor, and use it to perform on-chip absorption spectroscopy of \ce{CO2} concentrations down to \SI{0.1}{\percent} at $ \lambda\mkern-1mu =\mkern-1mu $~\SI{4.24}{\micro\meter} wavelength. We demonstrate that the $ \Gamma $, and hence the sensitivity to \ce{CO2}, of our waveguide is \SI{44}{\percent} that of free-space sensing.

%Our photonic waveguide for \ce{CO2} sensing, shown in Fig.~\ref{fig:device_setup}~(a)-(c), is a partially suspended \ce{Si} beam, supported by tapered \ce{SiO2} pillars, at a height of \SI{3}{\micro\metre} from the \ce{Si} handle substrate. 
Our photonic waveguide, shown in Fig.~\ref{fig:device_setup}~(a--c), is a \ce{Si} beam partially suspended \SI{3}{\micro\metre} above the \ce{Si} handle substrate and supported by tapered \ce{SiO2} pillars. 
The waveguide is \SI{220}{\nano\metre} thick and \SI{2.2}{\micro\metre} wide in its suspended sections.
At the pillars, spaced \SI{83}{\micro\meter} from each other, it widens to \SI{9}{\micro\meter} with \SI{17}{\micro\meter}-long linear tapers, i.e. \ang{11.5} tapers. According to bidirectional eigenmode expansion simulations (see Supplementary Figure~\ref{fig:vis1_anchor}), the support structures have an insertion loss of \SI{0.025}{\decibel} each, and are thus the main contributor to the waveguide propagation loss.
The absence of a continuous solid bottom cladding and the large separation from the substrate limit the substrate leakage loss to a simulated \SI{0.12}{\decibel/\centi\meter} and make the waveguide vertically symmetric, thus allowing light guiding with a very small core thickness, and hence low confinement of the propagating light. These features result in a high $ \Gamma $, negligible material absorption and substrate leakage losses, and maximized mode overlap with the analyte gas in all directions.
According to finite-element-method (FEM) simulations, the supported quasi-TE fundamental mode at \SI{4.24}{\micro\metre} wavelength (Fig.~\ref{fig:device_setup}~(d)) has an EFR of \SI{65.4}{\percent} and an external confinement factor $ \Gamma $ of \SI{52.0}{\percent}.
Taking into account the support structures, the simulated effective EFR and $ \Gamma $ are \SI{63.2}{\percent} and \SI{46.3}{\percent}, respectively. We note that the high value of $ \Gamma $ results from the low confinement of the light and not from narrow-band effects such as slow light enhancement or resonance.   
Furthermore, the waveguide features a small footprint and can be routed to form photonic circuits. We designed a waveguide circuit layout, shown in Fig.~\ref{fig:device_setup}~(e), for fully on-chip-referenced measurements, to avoid characterization errors due to the \ce{CO2} in ambient air. After an edge-coupled input waveguide section (see Supplementary Figure~\ref{fig:vis3_grating}~(a)), three outputs branch out with symmetrical 1x2 multi-mode interference (MMI) splitters at regular length intervals of \SI{1.6}{\milli\meter}, and terminate with surface grating couplers.
The MMI splitters are rectangular, non-tapered, \SI{8.4}{\micro\meter} wide and \SI{15}{\micro\meter} long (see Supplementary Figure~\ref{fig:vis2_mmi}). Fabrication error resulted in a residual \ce{SiO2} pillar underneath each MMI and rough sidewalls, which add a significant loss.
The grating couplers are fully suspended, through etched, and apodized to minimize backward reflection (see Supplementary Figure~\ref{fig:vis3_grating}~(b--d)). Their simulated back-reflection is \SI{5}{\percent}, and the upward radiation efficiency is \SI{45}{\percent}.      

\begin{figure*}[h!]
	%	\centering\includegraphics[width=\textwidth]{figures/GRoutput_and_fitloss_figure}
	\centering\includegraphics[width=\textwidth]{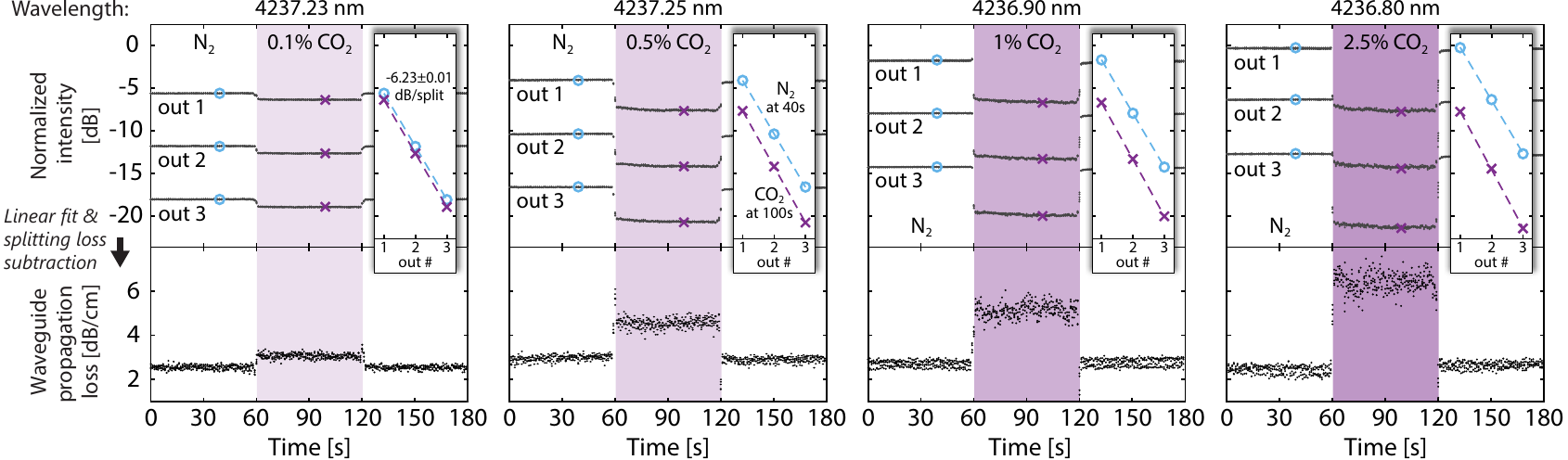}
	\caption{Four sample measurements of \SI{0.1}{\percent}, \SI{0.5}{\percent}, \SI{1}{\percent}, and \SI{2.5}{\percent} \ce{CO2} in \ce{N2}. The top panels show the waveguide output intensities, while the bottom panels show the extracted propagation loss. The insets show the linear fit of the intensities for two sample frames.}
	\label{fig:fitloss}
\end{figure*}

The waveguide circuit was fabricated on a commercial silicon-on-insulator (SOI) substrate with a \SI{220}{\nano\metre} \ce{Si} device layer and a \SI{3}{\micro\metre} \ce{SiO2} buried oxide (BOX) layer by a single electron-beam lithography step, dry etching of the \ce{Si}, wet etching of the \ce{SiO2}, and cleaving. We used electron-beam lithography because of its rapid turnaround time, but the minimum feature size in the circuit is compatible with stepper photolithography.

We characterized the \ce{CO2} sensing performance of our photonic waveguide with the setup shown in Fig.~\ref{fig:device_setup}~(e).
We focused \SI{4.24}{\micro\meter} continuous-wave linearly polarized light from a distributed-feedback quantum cascade tunable laser (MLQD4232, Thorlabs, USA) with single-wavelength emission onto the input facet of the waveguide. We placed the waveguide chip inside a steel case with a mid-IR-transparent window that allowed the visualization of the chip surface by a mid-IR camera (A6700sc, FLIR, USA) equipped with a cooled \ce{InSb} detector and a 1$\times$ macro lens. The chip case also had an inlet and an outlet that enabled controlled gas injection and a steady flow inside the case. The mid-IR camera aided the alignment of the waveguide input to the focused light for in-coupling and detected the output signal from the grating couplers.
We alternately injected nitrogen (\ce{N2}) and dilutions of \ce{CO2} in \ce{N2}, purchased pre-mixed, in concentrations of \SI{0.1}{\percent}, \SI{0.5}{\percent}, \SI{1}{\percent}, and \SI{2.5}{\percent} in \SI{1}{\minute} intervals at a flow rate of \SI{600}{\milli\liter/\minute}.
%We alternated a \SI{600}{\milli\liter/\minute} flow of nitrogen (\ce{N2}) and dilutions of \ce{CO2} in \ce{N2}, purchased pre-mixed, in concentrations of \SI{0.1}{\percent}, \SI{0.5}{\percent}, \SI{1}{\percent}, and \SI{2.5}{\percent} in \SI{1}{\minute} intervals.
For each \ce{CO2} concentration, we repeated the three-minute \ce{N2}-\ce{CO2}-\ce{N2} measurement at different wavelengths, calibrated with an accuracy of $\pm$~\SI{0.01}{\nano\meter}, across the \ce{CO2} absorption peak highlighted in the inset of Fig.~\ref{fig:atm_absorpt}. 
%The wavelength was calibrated exploiting the fixed position of the \ce{CO2} absorption line with an accuracy of $\pm$\SI{0.01}{\nano\meter}.
To reduce light absorption by the atmospheric \ce{CO2} along the free-space path between the laser head and the focusing lens and between the chip case and the camera, we enclosed these sections in brass tubes and continuously flushed those with \ce{N2}.

Fig.~\ref{fig:fitloss} shows the results of four sample measurements, one for each \ce{CO2} concentration tested. The top panels show the intensity time traces of the three waveguide outputs, including all recorded frames. 
Here, the intensity drop during \ce{CO2} injection results from light absorption along the entire light path, i.e.\ along the waveguide, including the input section before the first output, and the free-space sections in the setup, where residual ambient \ce{CO2} is present despite the \ce{N2} flushing.
By extracting the decay rate of the light intensity along the waveguide from the three outputs, as shown in the insets of Fig.~\ref{fig:fitloss}, we continuously measured the waveguide propagation loss in a real-time cut-back measurement. We accounted for the splitting loss, measured using a reference waveguide circuit with no additional waveguide length between splitters (Fig.~\ref{fig:device_setup}~(e), Supplementary Figure~\ref{fig:vis2_mmi}) to be \SI{5.8(2)}{\decibel/split}. 
%(including the \SI{3}{\decibel/split} due to the power halving).
%\SI[multi-part-units = single]{6.05(59)}{\decibel}
%The measured propagation loss, displayed in the bottom panels of Fig.~\ref{fig:fitloss}, excludes the absorption effects external to the straight waveguide sections between grating outputs.
\begin{figure}[t]
	%	\centering\includegraphics[width=.8\textwidth]{figures/CO2_loss_figure_wl-hit}
	%	\centering\includegraphics[width=.8\textwidth]{figures/CO2_loss_figure_wl-hit_1co2col}
	\centering\includegraphics[width=\linewidth]{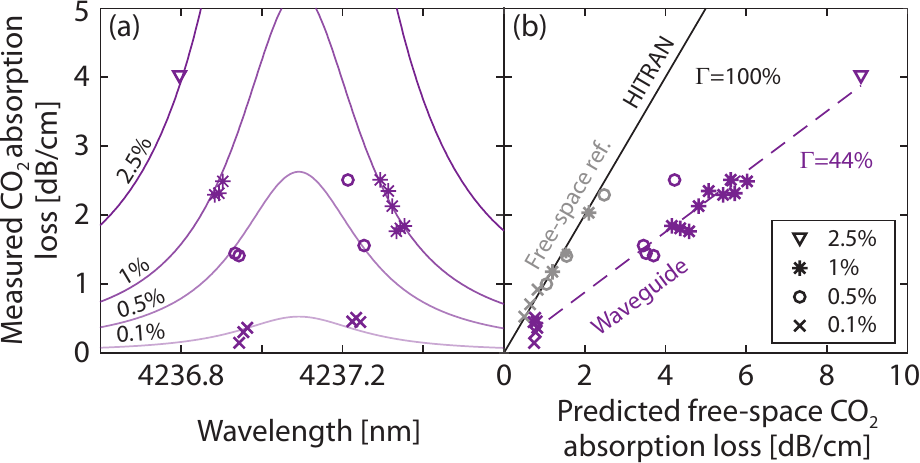}
	\caption{Measured excess waveguide propagation loss due to \ce{CO2}, compared to the free-space \ce{CO2} absorption loss~\cite{Gordon2017hitran2016}. (a)~Waveguide \ce{CO2} loss spectrum (markers) and \SI{44}{\percent}-downscaled HITRAN free-space spectrum (solid lines). (b)~Waveguide \ce{CO2} loss as a function of the free-space loss. The gray markers show our free-space reference measurement.}
	\label{fig:CO2loss}
\end{figure} 
The measured propagation loss is displayed in the bottom panels of Fig.~\ref{fig:fitloss}.
%The propagation loss during \ce{N2} injection, i.e.\ the waveguide base loss $ \alpha_{dB}\mkern-1mu =\mkern-1mu \frac{10}{\ln(10)}\alpha $, is \SI{4.6}{\decibel/\centi\meter}. 
The propagation loss during \ce{N2} injection, i.e.\ the waveguide base loss, is \SI{3(1)}{\decibel/\centi\meter}. 
%The difference between the time-averaged waveguide propagation loss during \ce{N2} injection and the one during \ce{CO2} injection is the waveguide excess propagation loss due to \ce{CO2}, i.e.\ the waveguide \ce{CO2} absorption loss.
The increase in the loss during \ce{CO2} injection, clearly visible at all tested concentrations, is the excess loss caused exclusively by absorption along the straight waveguide sections between grating outputs.
The response time is less than \SI{2}{\second} and limited by the manual gas switching and by the gas exchange time in the chip case.
The difference between the time-averaged propagation loss during \ce{N2} injection and the one during \ce{CO2} injection is the waveguide \ce{CO2} absorption loss.

Fig.~\ref{fig:CO2loss} shows the waveguide \ce{CO2} absorption loss for all concentrations and wavelengths measured.  
We compare the measured waveguide \ce{CO2} absorption loss with the predicted free-space \ce{CO2} absorption loss at corresponding pressure and temperature, as listed in the HITRAN database~\cite{Gordon2017hitran2016} and confirmed by our free-space reference measurement. We find that the $ \Gamma $ of our waveguide, i.e.\ its sensitivity to \ce{CO2}, is \SI{43.85(01)}{\percent} that of free-space, a value close to the simulated \SI{46.3}{\percent}.
%\SI[multi-part-units=repeat]{43.85(01)}{\percent}
The measured FOM of our waveguide is thus \SI{0.6(3)}{\centi\meter}.
The standard deviation of the measured waveguide propagation loss in \ce{N2} (Fig.~\ref{fig:fitloss}) indicates that the smallest loss change measurable in our setup is \SI{0.18}{\decibel}, corresponding to a \ce{CO2} concentration change of \SI{350}{ppm}.
According to the measured FOM, the optimal length~\cite{Kita2018are,Siebert2005infrareda} for our waveguide to sense present-day atmospheric levels of \ce{CO2}, i.e. 400 ppm, is \SI{1.4}{\centi\meter}.     

In Table~\ref{tab:comparison}, we compare our waveguide to other relevant integrated gas-sensing waveguides.
Our waveguide features the highest theoretical and experimentally demonstrated $ \Gamma $ and FOM amongst all the listed waveguides.
%Our waveguide features the highest theoretical and experimentally demonstrated $ \Gamma $ amongst all the listed waveguides, and exhibits the highest FOM amongst the mid-IR waveguides. 
%Our waveguide design features the highest theoretically achievable $ \Gamma $ and FOM amongst all the listed waveguides. It also exhibits the highest experimentally demonstrated $ \Gamma $ of all, and the highest FOM amongst the mid-IR waveguides.  

We note that Ranacher et al.~\cite{Ranacher2018characterization,Ranacher2018midinfrared} measured a $ \Gamma $ consistently higher than the simulated one. This might be caused by the absorption and subsequent release of \ce{CO2} by the plastic tubing and chip case used in the experiments. Such memory effects are particularly relevant when injecting the \ce{CO2} mixtures in order of decreasing concentration, and result in higher \ce{CO2} levels than intended. Ultimately, the performance of these waveguides is limited by the high intrinsic mode loss caused by the mid-IR absorption of the \ce{SiO2} cladding and large support structures. \ce{SiO2} absorption at \SI{4.23}{\micro\meter} wavelength is, in fact, \SI{54}{\decibel/\centi\meter}~\cite{Kischkat2012midinfrared}.

Tombez et al.~\cite{Tombez2017methane} achieved a high FOM in methane (\ce{CH4}) sensing by probing an overtone absorption band of \ce{CH4} using the fundamental TM-polarized mode at \SI{1.65}{\micro\meter} wavelength. For their waveguide design, and theoretically for all the listed waveguides, the TM mode is less confined than the TE mode, thus more sensitive, but also lossier. Tombez et al.\ can exploit the TM mode while keeping the waveguide base loss low because they employ a wavelength at which the \ce{SiO2} bottom cladding features low absorption. Such strategy does not work at longer wavelengths, at which the TM mode's base loss increases dramatically. 
Additionally, the mode size increases with the wavelength, and requires increasing the spacing between the waveguide and a high-refractive-index substrate to limit substrate leakage loss.
%as indicated by Han et al.'s waveguide~\cite{Han2016onchip}.

We achieve a high $ \Gamma $ with the fundamental TE mode thanks to the small waveguide thickness, and a potentially low base loss thanks to the almost complete removal of the \SI{3}{\micro\meter}-thick \ce{SiO2} BOX layer.
The intrinsic loss could be lowered by optimizing the support structure design and increasing their pitch.  
Furthermore, a higher-quality fabrication process, possibly including thermal oxidation and selective oxide etching to smoothen the waveguide surfaces, could further reduce the loss. Since the etched sidewalls constitute only one tenth of the waveguide surface, our design may achieve very low scattering losses.
%The discrepancy between the simulated and the measured base loss is due the uncertainty of the splitting loss measurement and of the of the simulation, which relies on an approximated model of the support structures.
%The difference between the simulated and the measured base loss is due to sidewall roughness.   
%The loss could be reduced with a higher-quality fabrication process, possibly including thermal oxidation and selective oxide etching to smoothen the waveguide surfaces. Since the etched sidewalls constitute less than one tenth of the waveguide surface, our design may achieve very low scattering losses. Moreover, optimization of the support structure design and an increase in their pitch could further reduce the intrinsic loss. 

\begin{table}[t]
	\centering
	\caption{Comparison of reported gas-sensing waveguides.}
	\resizebox{\columnwidth}{!}{
		%	\resizebox{.7\textwidth}{!}{
		\begin{threeparttable}
			%			\caption{Table}
			%			\small\addtolength{\tabcolsep}{-2pt}
			\addtolength{\tabcolsep}{-2pt}
			\begin{tabular}{l l c c c c c c c c} 
				\toprule
				& & & & \multicolumn{3}{c}{Simulated} &  \multicolumn{3}{c}{Measured} \\
				\cmidrule(r){5-7} \cmidrule(r){8-10}
				& Pol. & \makecell{$ \lambda $\\{}[\si{\micro\meter}]} & Gas & \makecell{Intrinsic\\mode loss\tnote{*}\\ {}[\si{\decibel/\centi\meter}]}  & \makecell{$\Gamma$\\ {}[\si{\percent}]} & \makecell{FOM\\$ \nicefrac{\Gamma}{\alpha} $\\ {}[\si{cm}]} & \makecell{Waveguide\\base loss $ \alpha_{dB} $\\ {}[\si{\decibel/\centi\meter}]} & \makecell{$\Gamma$\\ {}[\si{\percent}]} & \makecell{FOM\\$ \nicefrac{\Gamma}{\alpha} $\\ {}[\si{cm}]} \\
				\midrule
				%				Susp wg with SUBSTRATE wgw2.1: Mode loss = 0.16 dB/cm, Gamma = 54.49%, EFR = 				65.67% 
				%				Susp wg with SUBSTRATE wgw2.2: Mode loss = 0.12 dB/cm, Gamma = 52.01%, EFR = 				65.40%
				\makecell[l]{This\\work} & TE & 4.24 & \ce{CO2} & \num{2.9} & \num{46.3} & \num{0.69} & \num{3(1)} & \num{44} & \num{0.6(3)} \\
				\cite{Ranacher2018characterization} & TE & 4.23 & \ce{CO2} & \num{4.4} & \num{10.8} & \num{0.11} & \num{4.0}\tnote{$\ddagger$} & \num{14} & \num{0.15} \\
				\cite{Ranacher2018midinfrared} & TE & 4.23 & \ce{CO2} & \num{7.6} & \num{11.6} & \num{0.07} & - & \num{19.5} & - \\
				%				Ranacher et al.~\cite{Ranacher2018characterization}, TE & 4.23 & \num{8.6} & \num{4.66} & \num{11} & \num{0.056} & \multirow{2}{*}{\num{3.98}} & \multirow{2}{*}{\num{14}} & \multirow{2}{*}{\num{0.083}}  \\
				%				Ranacher et al.~\cite{Ranacher2018characterization}, TM & 4.23 & \num{11.5} & \num{47.96} & \num{29} & \num{0.014} & & & \\
				%				Ranacher et al.~\cite{Ranacher2018midinfrared}, TE & 4.23 & \num{12} & \num{0.35} & \num{13.1} & \num{1.024} & \multirow{2}{*}{not reported} & \multirow{2}{*}{\num{19.5}} & \multirow{2}{*}{-} \\
				%				Ranacher et al.~\cite{Ranacher2018midinfrared}, TM & 4.23 & \num{41.3} & \num{5.76} & \num{105.7} & \num{0.434} & & & \\
				\cite{Tombez2017methane} & TM & \num{1.65} & \ce{CH4} & \num{1.7} & \num{28.3} & \num{0.71} & \num{2} & \num{25.4} & \num{0.55} \\
				%				\cite{Han2016onchip} & \makecell[l]{TE\\TM} & \num{3.31} & \ce{CH4} & \makecell{\num{1.34}\\ \num{4.67}} & \makecell{\num{4.60}\\ \num{8.93}} & \makecell{\num{0.15}\\ \num{0.08}} & \num{7} & \num{2.3} & \num{0.01} \\
				%%				\cite{Han2016onchip} (TM) & & \num{5} & \num{10.3} & \num{0.062} & & & \\ 
				\bottomrule	
			\end{tabular}
			\begin{tablenotes}\footnotesize
				\item[*] Assuming zero gas absorption, excluding scattering losses, and including substrate leakage losses and, for the suspended designs, losses due to the support structures.
				%				\item[$\dagger$] Including the loss due to the support structures .
				%				\item[$\ddagger$] It accounts for the wider waveguide sections exposed to gas.
				%				\item[$\dagger$] Including substrate leakage loss and support structures loss. 
				\item[$\dagger$] Measured at \SI{4.17}{\micro\meter} wavelength.   
			\end{tablenotes}
		\end{threeparttable}
	}
	\label{tab:comparison}
\end{table}

In conclusion, we have demonstrated the optical absorption spectroscopy of \ce{CO2} concentrations down to \SI{0.1}{\percent} using a \SI{3.2}{\milli\meter}-long low-confinement \ce{Si} photonic waveguide for \SI{4.24}{\micro\meter} wavelength. The waveguide was fabricated on an SOI platform with a single lithography step. The waveguide, intrinsically broadband, can operate in conjunction with both broad- and narrow-band sources and detectors. It has a small footprint and can easily be routed to form mid-IR photonic circuits, potentially including components such as resonant cavities and spectral filters. By integrating MMI splitters, we implemented for the first time a waveguide circuit for on-chip-referenced gas measurements, to eliminate errors due to the ambient \ce{CO2} and ensure that the characterized sensing performance is ascribable only to the waveguide. In this way, we have demonstrated that the $ \Gamma $, and hence the sensitivity to gas, of our waveguide is \SI{44}{\percent} that of free-space sensing, and its FOM is \SI{0.6}{\centi\meter}. Compared to previous mid-IR gas-sensing waveguides, our waveguide exhibits the highest external confinement factor and a four-fold improved FOM. This demonstrated performance and the simple, cost-effective, and scalable fabrication make our integrated photonic waveguide ideal for mass production and large-scale adoption. It has the potential to become the choice component for an increasingly broad range of applications, such as portable and distributed environmental monitoring, and high-volume medical and consumer applications.

\section*{Funding Information}
This work was partially funded by grants from VINNOVA (2016-02328 and 2017-05108), SLL (20150910), and SSF (GMT14-0071).

%{\footnotesize
{\small
\bibliography{gassens_paper_mod}
}

\newpage

\setcounter{figure}{0}

\section*{Supplementary figures}

\begin{figure}[ht]
	\centering\includegraphics[width=\linewidth]{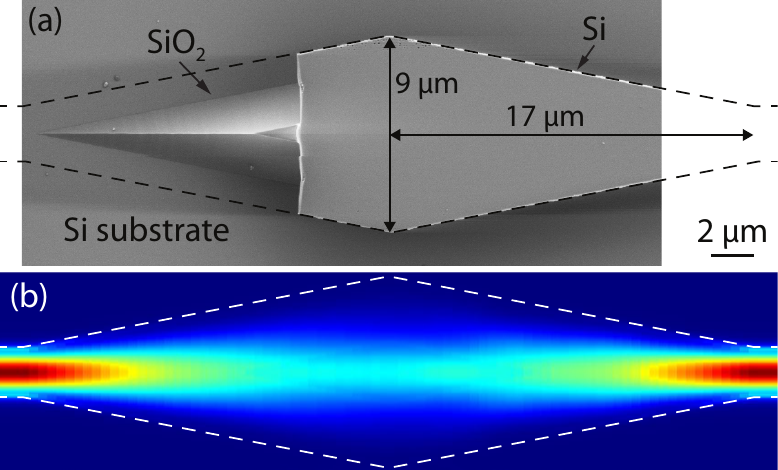}
	\caption{(a)~Top-view SEM image of a waveguide support structure where the Si waveguide was partially removed to expose the SiO2 pillar below. (b) Top view of the time-averaged squared magnitude of the electric field distribution in the middle plane of the waveguide, according to bidirectional eigenmode expansion simulations. The simulated insertion loss of the support structure is \SI{0.025}{\decibel}.}
	\label{fig:vis1_anchor}
\end{figure}

\medskip
\begin{figure}[h!]
	\centering\includegraphics[width=\linewidth]{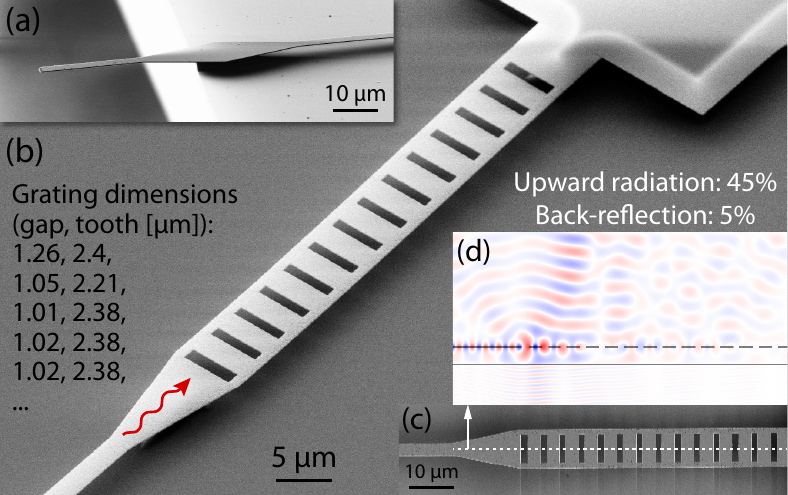}
	\caption{(a)~Perspective-view SEM image of the suspended waveguide input at the chip edge. (b)~Perspective and (c)~top SEM views of a suspended surface grating output coupler. The grating is through-etched and apodized. (d)~Propagating electric field in the vertical plane marked in (c), according to eigenmode expansion simulations.}
	\label{fig:vis3_grating}
\end{figure}

\newpage
\mbox{}
\medskip
\begin{figure}[ht]
	\centering\includegraphics[width=\linewidth]{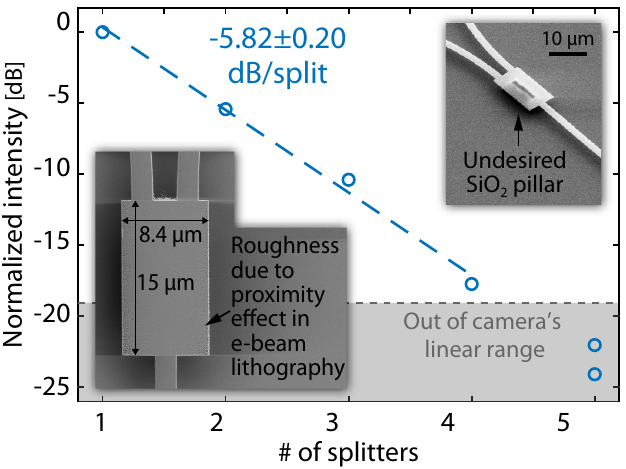}
	\caption{The measured MMI splitter loss is \SI{5.8(2)}{\decibel/split}. 
		%		The insets show SEM images of (top-right inset) a perspective view of the residual \ce{SiO2} pillar underneath an MMI splitter and (bottom-left inset) a top view of its rough sidewalls.
		The insets show a perspective-view (top-right inset) and a top-view (bottom-left inset) SEM images of an MMI splitter, where the residual \ce{SiO2} pillar underneath it and the rough sidewalls are visible, respectively.
		These defects resulted from fabrication error and add a significant loss.}
	\label{fig:vis2_mmi}
\end{figure}

\end{document}